\documentclass[12pt,journal,compsoc]{IEEEtran}

\pdfoutput=1

\usepackage[T1]{fontenc}
\usepackage{float}
\usepackage{graphicx}

\usepackage[]{footmisc}

\usepackage{multicol}

\newcommand{\code}[1]{\texttt{#1}}
\newcommand{\figref}[1]{Fig. \ref{fig:#1}}

\usepackage{listings}
\usepackage{color}

\definecolor{dkgreen}{rgb}{0,0.6,0}
\definecolor{gray}{rgb}{0.5,0.5,0.5}
\definecolor{mauve}{rgb}{0.58,0,0.82}

\begin{document}

\author{
    \IEEEauthorblockN{Abolfazl Danayi\IEEEauthorrefmark{1}, Saeed Sharifian\IEEEauthorrefmark{2}}\\
    \IEEEauthorblockA{\IEEEauthorrefmark{1}Amirkabir University of Technology, Tehran, Iran
    \\adanayidet@gmail.com}\\
    \IEEEauthorblockA{\IEEEauthorrefmark{2}Amirkabir University of Technology, Tehran, Iran
    \\sharifian\_s@aut.ac.ir}
}

\IEEEtitleabstractindextext{%
\begin{abstract}
In order to address the complexity and extensiveness of technology, Cloud Computing is utilized with four main service models. The most recent service model, function-as-a-service, enables developers to develop their application in a function-based structure and then deploy it to the Cloud. Using an optimum elastic auto-scaling, the performance of executing an application over FaaS Cloud, overcomes the extra overhead and reduces the total cost. However, researchers need a simple and well-documented FaaS Cloud manager in order to implement their proposed Auto-scaling algorithms. In this paper, we represent the openCoT platform and explain its building blocks and details. Experimental results show that executing  a function (invoking and passing arguments) and returning the result using openCoT takes 21 ms over a remote connection. The source code of openCoT is available in the GitHub repository of the project (\code{www.github.com/adanayi/opencot}) for public usage.
\end{abstract}

}

\title{openCoT: The opensource Cloud of Things platform}

\maketitle
\IEEEdisplaynontitleabstractindextext
\IEEEpeerreviewmaketitle

\section{Introduction}
Cloud Computing is one of the answers to the problem of the growing complexity and extensiveness of the technology. Besides, it provides a faster platform for application development and a better way for updating an application. Due to the definition of NIST in \cite{cloud:nist}, a cloud achieves this goal by providing on demand resource that can be rapidly provisioned and freed. In order to achieve this goal, clouds offer different service models. In both academia and enterprises, IaaS, PaaS and SaaS are well known. A recently born service model is FaaS in which the cloud allows a user to execute a code in the form of a function and the user does not face the complexity of managing, scheduling and execution of the code over underlying resources \cite{faas:usingFaaS}. 

In order to make use of the benefits of this model optimally, the programming architecture must be reviewed and maybe modified. Historically, the monolithic programming architecture has been the dominant choice for both application development and execution \cite{microservices:monolithic}. However, besides this approach, there have been alternatives. Microservices architecture, as a subset of the Service Oriented Architectures \cite{microservices:soa}, is one of the potential optimal choices for utilization of FaaS Clouds \cite{faas:sfaas, faas:usingFaaS, faas:serverlessIsMore}. However, service oriented programming has some overheads (such as API calls) in comparison to plain programs and primarily it seems to suffer from lower efficiency. But when we look into the problem considering its execution on Clouds the scaling and resource allocation may result in a better efficiency \cite{serverless:QModel, faas:usingFaaS, serverless:application:disaster}. The usage of FaaS clouds for microservices programming is getting more attention and in \cite{faas:ubench}, authors have proposed a benchmark procedure for evaluation of the performance of FaaS clouds. 

In the cases of IaaS, PaaS and SaaS, researchers have proposed many papers on the Auto-scaling subject and this problem has received enough attention. However, as the FaaS is a recent service model, its Auto-scaling algorithms need a new branch of research work.
In this paper, we propose the openCoT platform which is designed and implemented in order to help researchers implement their cloud provisioning algorithms and analyze the output in the real world. openCoT has a modular design, and can be setup easily in a local or remote network.

The rest of this paper is structured as follows. In section \ref{sec:mainBlocks}, the main blocks of openCoT are defined and explained. Then, the underlying structure, abstracted as \textit{Node}, is covered in section \ref{sec:node}. In the next section, the heart of the system, \textit{Controller} module, and also the mechanisms which are used to connect \textit{Nodes} to the \textit{Controller}, will be proposed. In section \ref{sec:implementation}, the implementation details and experimental results are given, and in the last section conclusion and future works are established.

\section{Main blocks of the architecture} \label{sec:mainBlocks}
The main concepts in openCoT are Controller, Node and Function and Cloud Broker which is not a part of openCoT but is the external layer that uses the Controller and is considered as a part of the architecture. In this section, we introduce each building block using a top-down approach.  

\subsection{Cloud Broker}
The Broker uses Controller APIs in order to make use of the openCoT. Broker is responsible for:
\begin{itemize}
	\item Collecting user requests
	\item Formatting requests and inserting them into the openCoT
	\item Collecting returned values of function executions and passing them to corresponding user/application(s).
	\item Auto-scaling system using the openCoT's scaling API (The auto-scaling format will be discussed layer).
	\item Setting up ports table (Ports and communication mechanisms will be discussed later)
	\item Setting up initial state of the system
	\item Setting up a folder for functions source codes and introducing its path to the Controller
	\item Introducing the number of Nodes inside each cluster to openCoT
\end{itemize}

\subsection{Controller}
The heart of openCoT is the \textit{Controller}; The core that setups servers so that nodes can communicate with and on the other hand, provides a simple API for the Cloud Broker. Controller dispatches Function Execution Requests (FER) between nodes and scales the system based on the Broker's order.

\subsection{Nodes and Clusters}
A Node is a host computer that is setup and running the openCoT's \code{Node.py} program. A Node only needs to know the Internet (IP and port) address of the Clerk server (in Controller), and then it automatically starts pairing with the Controller, auto-scaling itself and also receiving Function Execution Requests alongside with executing them and returning the RET values. In order to support heterogeneity, we have also defined a concept called Cluster. A Cluster is a number of nodes with similar physical attributes. Nodes, automatically download the source codes of functions from the Controller and build Docker \cite{docker} images. On a Node there is a number of Function Execution Units. Each Function Execution Unit is a docker container that can run its corresponding function when invoked.

\subsection{Function, FER and RET}
\paragraph{Function}
A function is the only entity that the cloud's developer user has to give to the system. A function in openCoT is a standard python function that receives the FER and returns the RET. In the first release of openCoT the Python language \cite{python} is supported. Every function consists of a \code{func.py} and a \code{requirements.txt} file, inside a folder whose name is similar to the name of that function. The \code{func.py} file is the main code part of the function. The following script is the simplest example of a \code{func.py} file and explains the fixed structure of naming and definition of a function.

\begin{lstlisting}
def f(FER):
	return {'ret':'Hello Cloud of Things!'}
\end{lstlisting}

\paragraph{FER} The Function Execution Request (FER) is the input to the function and has a fixed structure as shown in the following script. The \code{INPUTS} is a dictionary and has an arbitrary structure which is defined by the developer user and will be announced to consuming users if needed. \code{METADATA} is the data provided by Cloud Broker (the upper layer of openCoT) and will be passed to the function too. The Cloud Broker should announce the structure of \code{METADATA} to the users. Please note that the FER is a defined entity and exists inside openCoT. In other words, each request for execution of a function shapes an FER and exists until a node (that can handle that function) receives it for execution. Cloud Broker has to provide all three fields when \textit{submitting} an FER to the system, tough the ID field will be eliminated from the FER when passing to the function for execution, and will be attached to the corresponding RET again and passed by Broker at the end. Thus, the broker knows the RET goes back to which request.

\begin{lstlisting}
{'id':ID, 'x':INPUTS, 'm':METADATA}
\end{lstlisting}

\paragraph{RET} RET, is the returned value and is a dictionary. RETs are defined and known entities in the openCoT, too; and they exist inside the system until be pulled by the broker. As mentioned before, When a function returns the dictionary value, the \code{ID} will be attached to it and will be returned to the Broker. RET values are in form of the following structure where \code{RET\_VAL} is the returned dictionary of the function and \code{stat} holds the status of the execution (i.g. \code{OK} and \code{ERROR}).

\begin{lstlisting}
{'id':ID, 'ret':{'stat':STATUS, 'val':RET_VAL}}
\end{lstlisting}


\section{FEU and Node} \label{sec:node}
\begin{figure}
\begin{center}
	\includegraphics[scale=0.5]{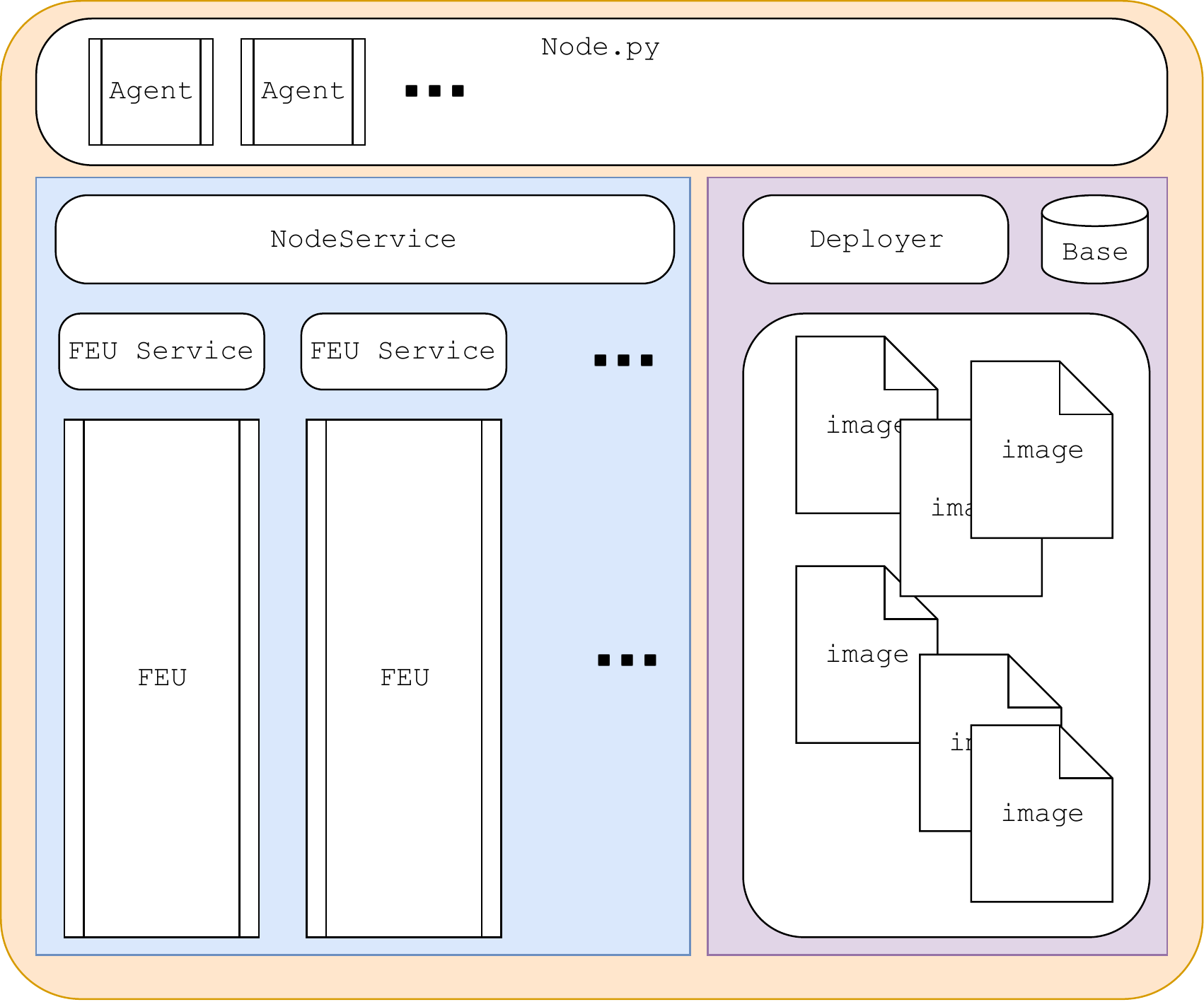}
	\caption{The structure of Node}
	\label{fig:node}
\end{center}
\end{figure}

In the previous section, we briefly explained Nodes and Clusters. In this section, we provide detailed information about Nodes. As mentioned before, on a Node there is a number of Function Execution Units (FEU) that run functions when invoked. Each FEU has its own docker \textbf{container} and is able to execute one FER in parallel with other FEUs. The structure of Nodes in openCoT is shown in \figref{node}. A node provides five characteristics:
\begin{itemize}
	\item Communications with the (remote) Controller
	\item Performing the Auto-scaling mechanism
	\item Performing the FER execution mechanism
	\item Performing the RET collection mechanism
	\item Performing the Container deployment mechanism
\end{itemize}

\subsection{Node, NodeService and Deployer}
As shown in \figref{node}, the three top classes are Node, NodeService and Deployer. \code{Node.py}, is the main building block and communicates with the Controller while the NodeService is responsible for creating, managing and communicating with FEUs via FEUService objects. However, when creating an FEU container, the corresponding Docker image must exist on the system. Thus, the Depolyer class is responsible for managing and creating FEU images when necessary. Alongside with FER execution and FEU image deployment, the third mechanism  which an openCoT Node provides is the Auto-scaling. These three mechanism are explained in the following subsections.

\subsection{Function execution mechanism: Agents, FEUService, FEU}
In the mentioned function execution mechanism shown in \figref{functions}, there are three modules involved in a Node. In this subsection we propose and explore each of them.

\subsubsection{Agent} On the Controller's side, a Gate is devoted for each function. A Gate consists of two entities: a Dispatcher which is the server that sends FERs to Nodes and also a Collector that receives RETs from Nodes named as PULL and PUSH servers, respectively. On the other side, Agents (on Nodes) are the Gate clients. An Agent is a \code{process} which asks the PUSH server for FERs. If there are no FERs submitted to the Gate, it waits a determined period of time, and then rechecks. In opposite, if the Gate responds with an FER, it schedules the FER to an available (non-busy) FEU via the NodeService and waits for the completion and then sends the results to the PULL server and restarts this cycle. It worth mentioning that the number of Agents for a function on a Node, is equal to the number of FEUs for that function. Thus, When an Agents asks NodeService for an available FEU, there is at least one available FEU; However, FEUs are not bound to Agents. In other words, an Agent can send its FER to any FEU.

\subsubsection{FEUService} This class is a \code{thread} that provides a between-process TCP/IP communication (Local-host) with FEUs and sends FERs to FEUs using this socket connection. It also receives RETs and passes them to the NodeService (and the Agent) using a python asynchronous scheme. FEUService is also responsible for sending the FIN message to the FEU which causes the FEU to finish its life-cycle and shutdown. Unlike Agents, FEUServices are bound to their corresponding FEUs.

\subsubsection{FEU} 
\begin{figure}
\begin{center}
	\includegraphics[scale=0.6]{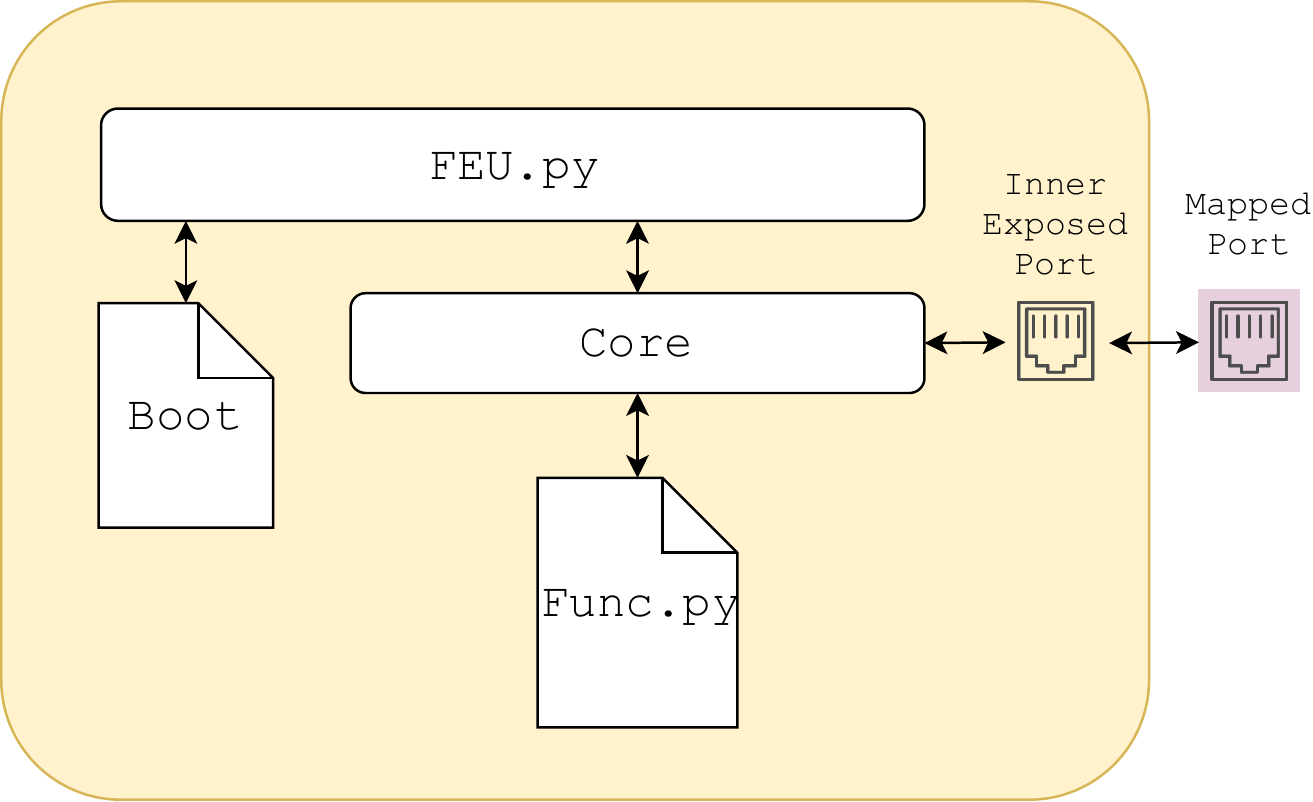}
	\caption{The structure of FEU}
	\label{fig:FEU}
\end{center}
\end{figure}

The Function Execution Unit (FEU) is a Docker based \code{Container} which receives FERs and passes them to the function which is implemented on it. The FEU structure is shown in \figref{FEU}. Each FEU Container immediately starts \code{FEU.py} program which reads the \code{Boot} file. The \code{Boot} file consists of the inner server's IP and Port. This server is implemented in the \code{Core.py} file, and listens to the dedicated Port. An important trick is that for all FEUs this inner port is the same within the container, but is mapped to a different outer port by the Docker engine and NodeService. In the first version of openCoT, \code{Core.py}'s server can handle two types of requests (primitives): \code{EXE} and \code{FIN}. When FEUService invokes the \code{EXE} primitive attached with the plain (bytes) FER, \code{Core.py} executes \code{Func.py} and returns the plain RET over TCP/IP connection to the FEUService. On the other hand, \code{FIN} primitive requests \code{Core.py} to close the program.

\subsection{Scaling mechanism}
Another important task of the Node class is the Scaling mechanism. As explained in the previous subsection, A node consists of FEUs, FEUServices and Agents and we also explained that on a Node, the numbers of FEUs, FEUServices and Agents are the same. We define the Scaling process of a Node as the process of allocating a Scaling Table over it. The transmission of this table from the Controller to the Node will be covered later in this paper. It is up to the NodeService class to perform the allocation. The structure of a Scaling Table is given in the following code segment. It is a dictionary where keys are the name of requested functions (same with FEU images), N$_i$ is the number of instances of that function and P$_i$ is the portion of CPU allocated to those N$_i$ FEUs. When NodeService receives a Scaling Table, it first flushes the Node, in other word, closes all Current FEUs by sending \code{FIN} primitive to them using their FEUServices. After that, the NodeService creates new FEUs and FEUServices and bounds each FEU to its corresponding FEUService; finally handles all created FEUServices to the upper layer, \code{Node.py}. Finally, \code{Node.py} creates $N_i$ agents for each function.

\begin{lstlisting}
{
	'function1':(N1, P1),
	'function2':(N2, P2),
	.
	.
	.
	'function_M':(N_M, P_M)
}
\end{lstlisting}

As an instance, the following Scaling Table requests for creation of 1 \code{hellocot} FEU with 10\% of the CPU, 2 \code{echo} FEUs with 10\% of the CPU and 3 \code{echo} FEUs 20\% of the CPU for each of them. 
\begin{lstlisting}
{
	'hellocot':(1, 0.1),
	'echo':(2, 0.1),
	'echo':(3, 0.2),
}
\end{lstlisting}

\subsection{Deployment mechanism}
Another key process is the Deployment mechanism. When \code{Node.py} receives the Scaling request (Scaling Table), before passing it to the NodeService, it checks if the images of requested FEUs exist (cached) on the Node. If one or more are missing, the \code{Node.py} requests the Clerk Server on the Controller for the source files, \code{func.py} and \code{requirements.txt}. Then using the \code{Deployer} class, creates FEU images for missing functions. We call this process Deployment. The Depolyer class as shown in \figref{node} has access to \code{Base} source files. The \code{Base} contains the following list:
\begin{itemize}
	\item \code{FEU.py} file
	\item \code{Core.py} file
	\item \code{Boot} file
	\item \code{common\_convs.py} file
	\item \code{Dockerfile}
\end{itemize}

The first three files are already known to the reader and, the same classes existing in the FEU. The \code{common\_convs.py} file is used inside the FEU but is not shown in \figref{FEU}. It is a simple library that provides dictionary-to-plainByte conversion functions which are used by FEUService and \code{Core.py}. The last file is the most important one, and is the file which has the settings for the docker engine in order to create the FEU image.

\section{Controller} \label{sec:controller}
\begin{figure}
\begin{center}
	\includegraphics[scale=0.55]{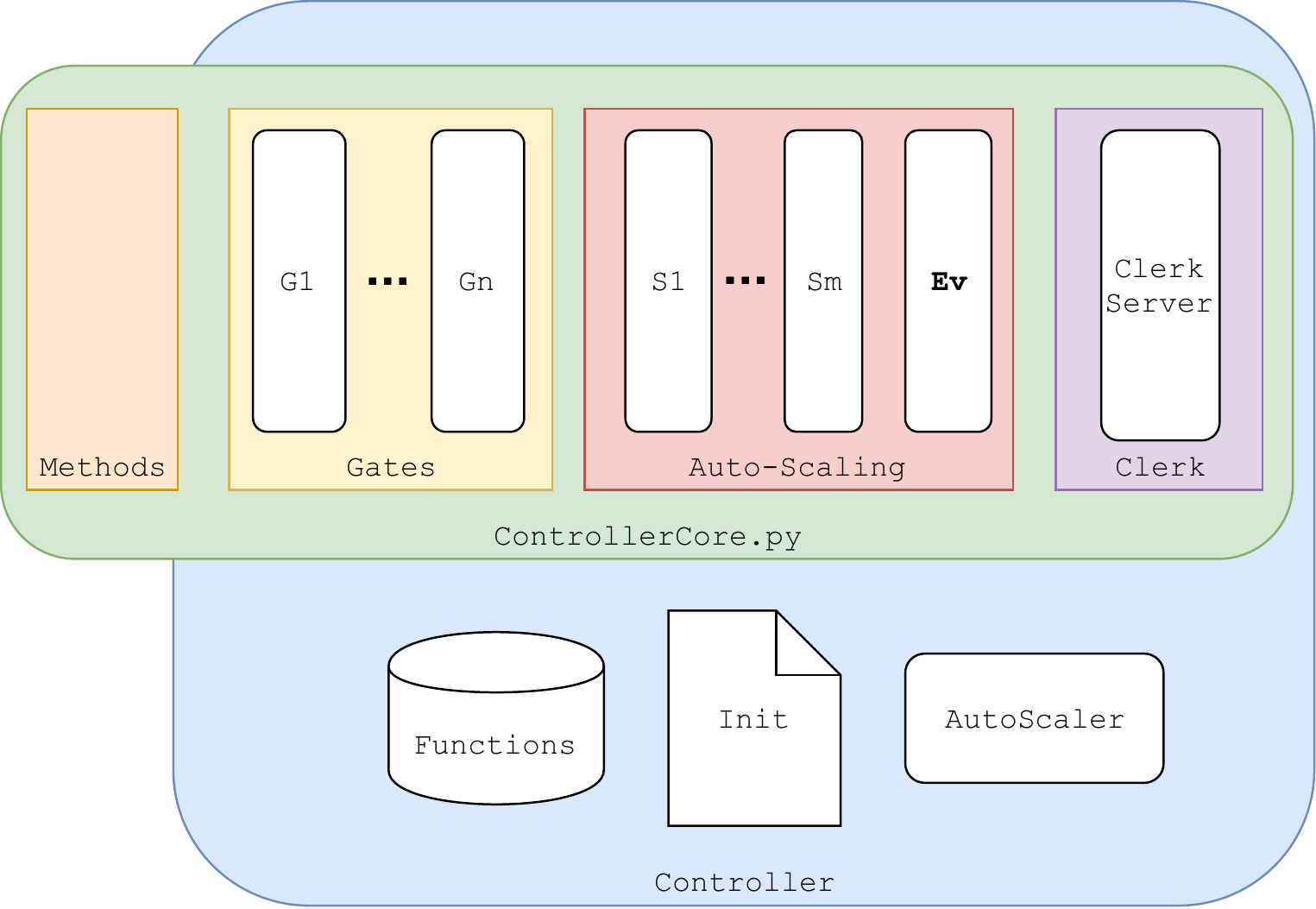}
	\caption{The structure of Controller}
	\label{fig:controller}
\end{center}
\end{figure}

The previous section explored the structure and functionalities of Nodes. In this section, we introduce the Controller and also the mechanisms which are used for Node-Controller communications that are build upon the high performance ZeroMQ (zmq) \cite{zmq}. The structure of the Controller is shown in \figref{controller}. Starting with the next subsection, we first explain this structure and then discuss the proposed communication mechanisms in each subsection.

\subsection{The structure of the Controller}
As shown in \figref{controller}, the Controller consists of four main communication mechanisms that are implemented in the \code{ControllerCore.py} class. We call each of these mentioned mechanisms a space. The first space, Methods space, is the wrapper of high level functions offered to the Cloud Broker. The Clerk server is used to announce the Internet address of other servers. Auto-scaling space consists of an Events server and a number of Scaling servers. And finally, the Gates space is responsible for sending tasks to Nodes. 

Furthermore, there are three more modules embedded in the Controller. The \code{functions} database is a root directory which consists of the source (\code{func.py}) and requirements files of each function within a sub-directory named with the label of that function. \code{Init} is also a directory that includes the initialization settings, such as ports table, the label of clusters and number of Nodes inside each cluster. The Autoscaler is another module that the \code{ControllerCore.py} utilizes in order to manage auto-scaling process and keep track of Nodes scaling.

\subsection{Methods space}
The Methods space provides high level methods (function calls) for the Cloud Broker. The main methods are listed and explained below.

\begin{itemize}
\item \textbf{Autoscale} This method receives the Auto-Scaling table. Although the structure of Node's Scaling table is explained, the Auto-Scaling table is a bit different. This table describes how many nodes within each cluster and with which Scaling table (can be more than one Scaling tables) are required.

\item \textbf{Push FER} The Push FER method receives the FER and the function label and pushes it in the FERs queue for that function label.

\item \textbf{Pop RET} The argument of this method is the function label and returns the RET object on a FIFO basis.

\item \textbf{Check Available} This method gets the function label and returns a True value if there are RET objects available in the queue of that function. The Broker can then call the \code{Pop RET} method to get this RET.
\end{itemize}

\subsection{Clerk Server}
The Clerk server is used by Nodes to query information from the Controller. As mentioned before we have used the ZeroMQ platform for communication. ZeroMQ offers four messaging patterns and the \code{REQ/REP} pattern fulfills the Clerk server's requirements better than other patterns. In this pattern, a node sends a \code{REQ} message and the the Clerk server replies (\code{REP}).

The first usage of the Clerk server is to query if the Controller is set or not. In this case, the Node sends a \code{chk} message and the server replies with a \code{OK} message. The second and the third types are the queries for Gate Ports table and Auto-Scaling Ports table. The forth case is the function source query in which the node asks the Clerk server for the source code of a function when it needs a function deployment.

\subsection{Auto-Scaling mechanism}
\begin{figure}
\begin{center}
	\includegraphics[scale=0.6]{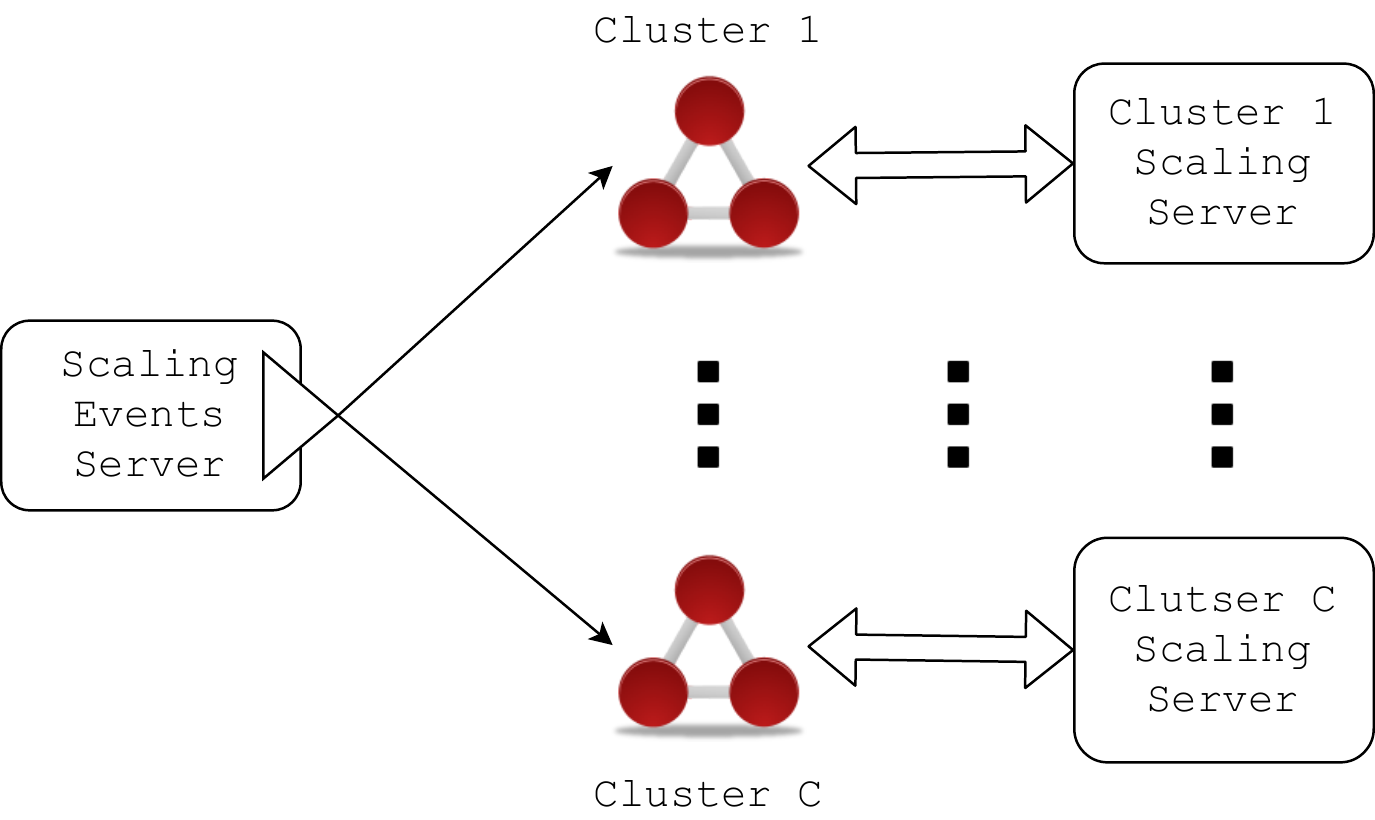}
	\caption{Auto-scaling space}
	\label{fig:autoscaling}
\end{center}
\end{figure}

As depicted in \figref{autoscaling}, for each cluster a Scaling server is dedicated. This server follows a \code{REQ/REP} pattern. In addition to this servers, the Scaling Events server is shared between all clusters and follows a \code{PUB/SUB} pattern. Whenever the Cloud broker submits a Auto-scaling table to the Cloud Broker, the Events server sends the Scaling event to all of the listening (Subscriber) nodes. As Nodes receive this event, they send a scaling request to the correlated Scaling server and the server returns the Scaling table to the node. It is also possible for the scaling server to return a \code{null} scaling table and in this case, the Node finds out that there is no need for its utilization and first flushes all of its working FEU, FEUService and agents and then scales itself out.

\subsection{Function execution mechanism}
\begin{figure}
\begin{center}
	\includegraphics[scale=0.42]{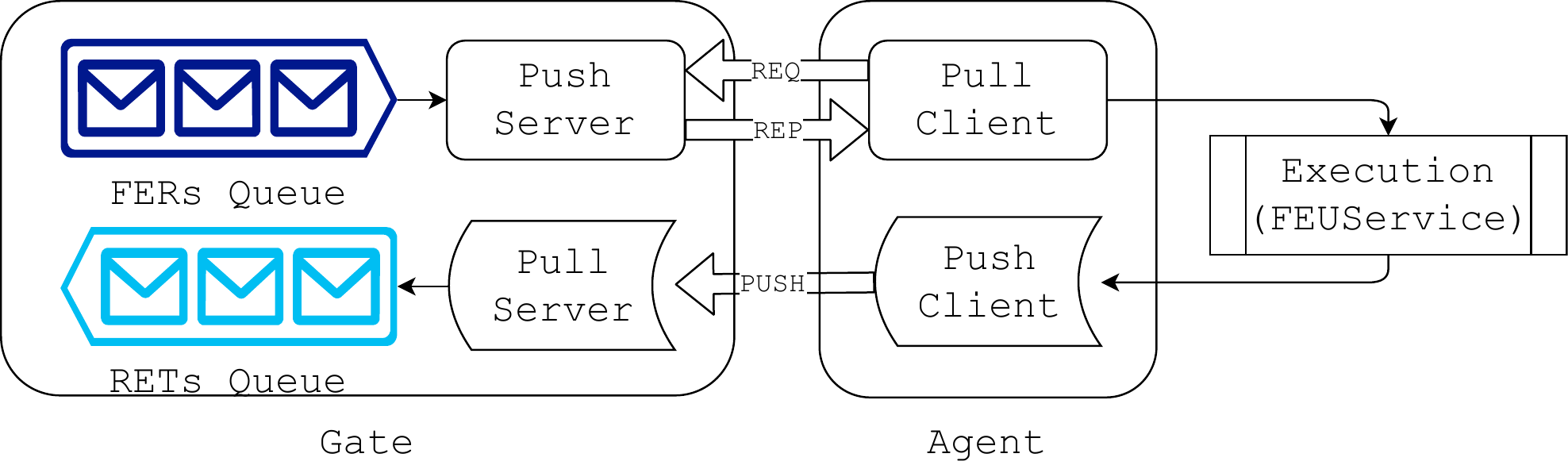}
	\caption{Function execution mechanism}
	\label{fig:functions}
\end{center}
\end{figure}

As mentioned before, the function execution mechanism has two main sides. A Gate which is devoted for each function on the Controller and also Agents on Nodes. This space is shown in \figref{functions}. When the Cloud Broker submits FERs using the Controller's methods space, the Controller pushes the FER into the FERs Queue of the related Gate. On the other hand, Agent's Pull client sends an \code{FER REQ} request to PUSH Server using a \code{REQ/REP} protocol. The Push server checks the FERs Queue and if there are FERs, pops them from the queue and sends them to the push client. In opposite when the queue is empty, push server returns a \code{Null} REP message. After the Push Server passes the FER to the Agent, it executes the FER and acquires the RET and then pushes the RET object to the Pull Server. The \code{PUSH/PULL} pattern is used for the connection between the Pull server and its clients.

\section{Experimental results} \label{sec:implementation}
In this section we check the performance and functionality of the openCoT in two scenarios. In Scenario-A, execution of the \code{hellocot}\footnote{Available as a default function in the Github repository of the project.} function is analyzed. In Scenario-B, openCoT is used to calculate the fast Fourier transform (FFT) of 100 blocks. Each block contains 256 samples. Two host computers are used in scenarios. $C_a$ has an Intel(R) Core(TM) i7-6500U CPU @ 2.50GHz with 8 GBytes of RAM memory and $C_b$ utilizes an AMD Athlon(tm) II X2 240 Processor @ 2.8 GHz with 3 GBytes of RAM. In both scenarios $C_a$ runs the Controller. It worth mentioning that the Controller computer can a Node too. In this case, the connection is made using the Local Host's IP\footnote{127.0.0.1}.

\subsection{Scenario-A}
In this scenario, first the \code{hellocot} function is executed in a typical Python environment 1000 times on the $C_a$ computer. Results show that the mean execution time is 1.196 microseconds with a standard deviation of 0.3 microseconds. 

\begin{figure}
\begin{center}
	\includegraphics[scale=0.55]{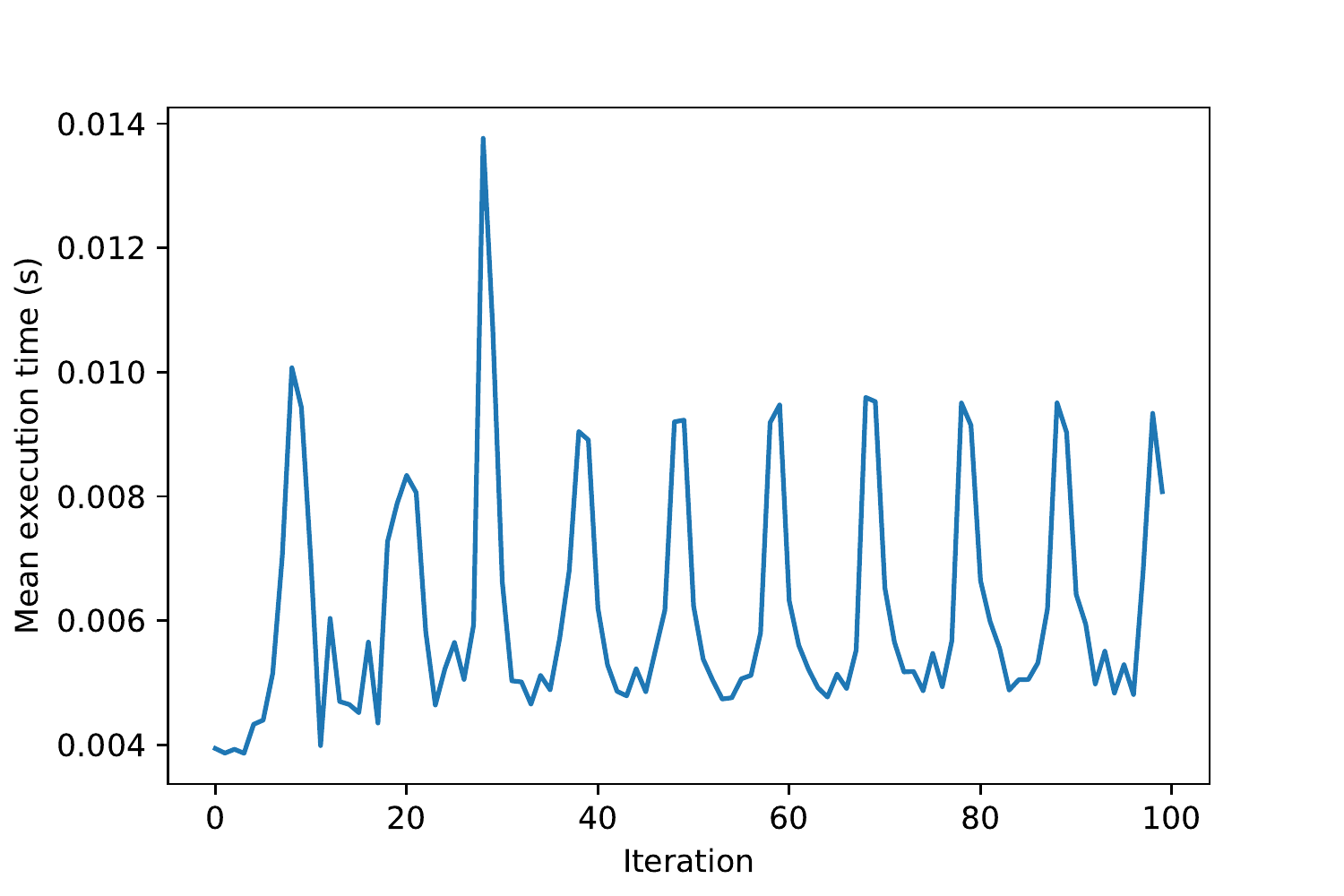}
	\caption{Scenario-A result 1}
	\label{fig:sc1}
\end{center}
\end{figure}

Then the $C_a$ is utilized to be the Controller and simultaneously a Node with 10 FEUs and this configuration is tested and the result is shown in \figref{sc1}. In each iteration 1000 FERs are submitted and when all RETs are received the mean execution time is calculated.

\begin{figure}
\begin{center}
	\includegraphics[scale=0.55]{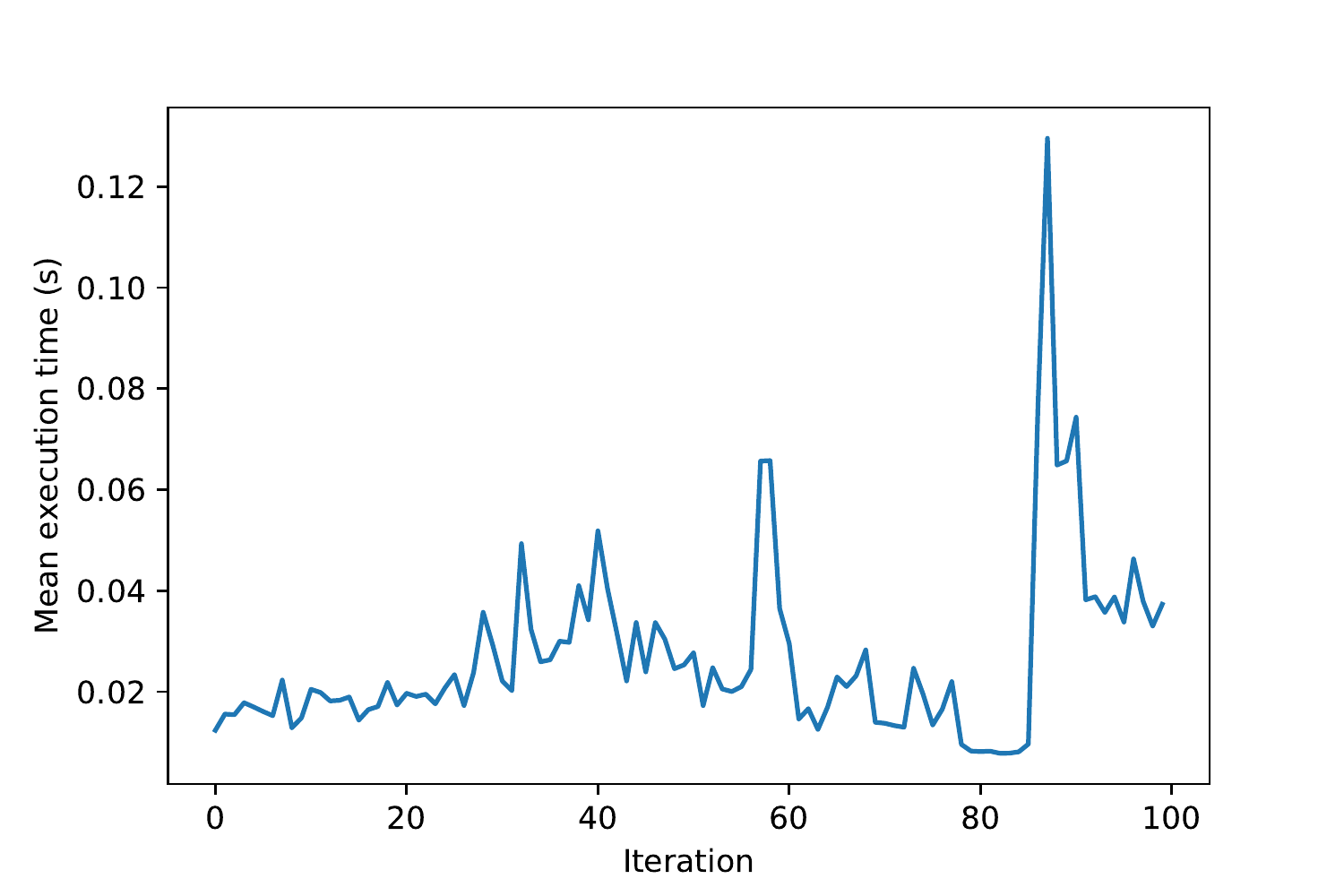}
	\caption{Scenario-A result 2}
	\label{fig:sc2}
\end{center}
\end{figure}

Finally, $C_b$ is used for the explained test, and the result is shown in \figref{sc2}.

\subsection{Scenario-B}
\begin{figure}
\begin{center}
	\includegraphics[scale=0.55]{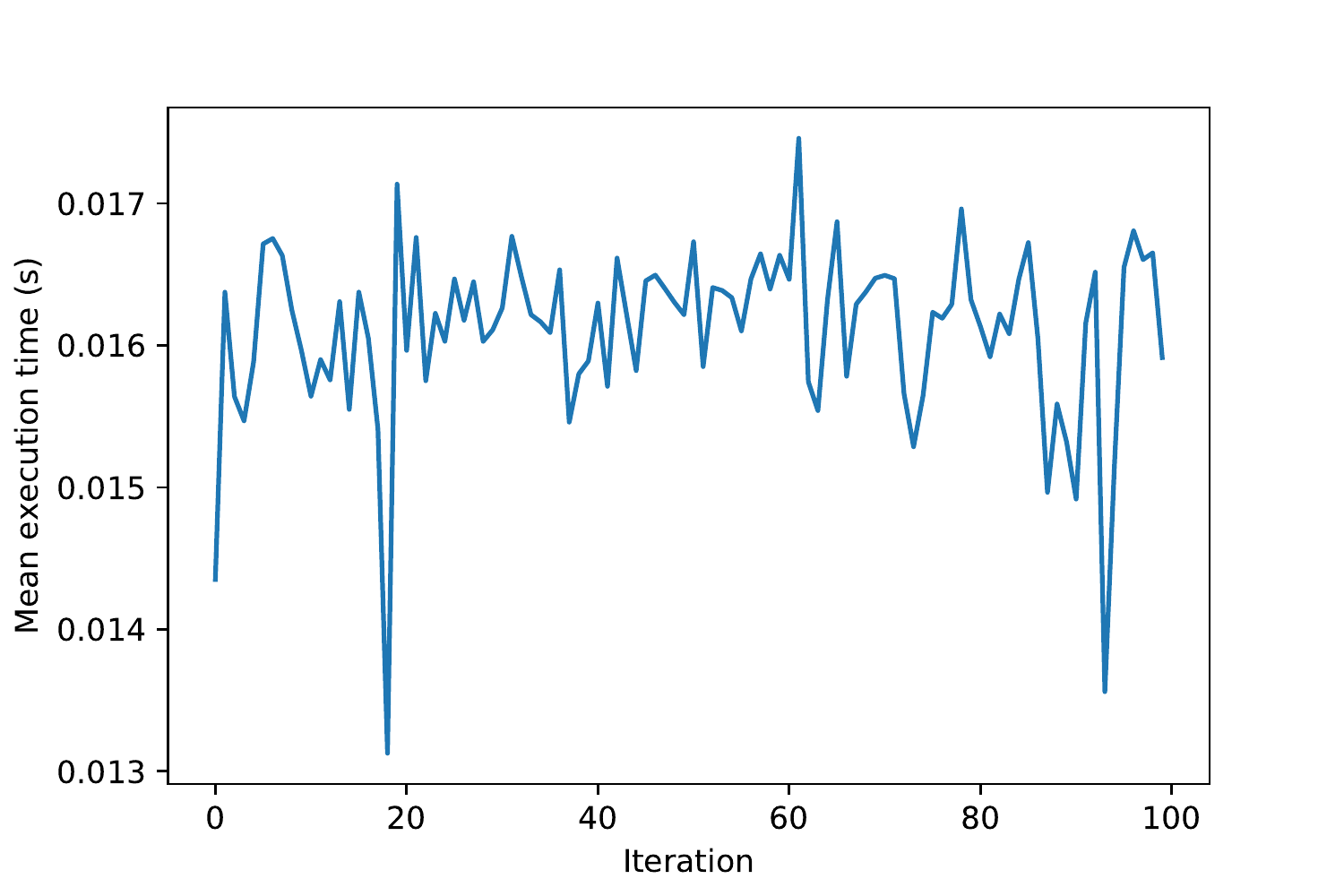}
	\caption{Scenario-B result}
	\label{fig:sc3}
\end{center}
\end{figure}

In this scenario a simple FFT function is implemented using the Numpy library \cite{numpy} and is again called 1000 times on $C_a$ outside openCoT for blocks of size 256. In this case we have 98 microseconds for the average execution time and 64 microseconds of standard deviation. Then, both $C_a$ and $C_b$ are used and the result is shown in \figref{sc3}.

Based on the result of Scenario-A1 the overhead of running a function using openCoT on the local host is 6.2 ms. Using the \code{ping} command, it is determined that the round-trip-time between $C_a$ is $C_b$ is 6.25 ms. Thus, the pure overhead in Scenario-A2, is 21 ms. The difference between these two values comes from the fact that $C_a$ is more powerful than $C_b$ and thus, it can be concluded that the overhead time is related to the power of the computer on both sides (Node and Controller).

According to the results of the Scenario-A, we can guess that the mean execution time of Scenario-B would be:
$$ \frac{10 . T_{C_a} + 10 . T_{C_b}}{10 + 10} = 13.6 s$$
However, the mean execution time is a bit more and is 16.1. The difference comes from the high data rate of the Scenario-B as the values are sent and received in the ASCII format.

\section{Conclusion and Future works} \label{sec:conclude}

In this work, motivated by the goal of enabling academia with a simple and well-documented FaaS Cloud platform for research purposes, we introduced the openCoT platform and then described its main blocks which are Broker, Controller, Node and FEU. After that, we explained the structure of Node and FEUs in details. Three communication mechanisms were proposed for the Controller-Node connection. We tested the performance of openCoT and measured the overhead time added to the function execution. Experimental results show that on a Local Host, 6.25 ms is added by the openCoT, where it is increased to 21 ms over remote connection. Based on these results, we suggest the following list as the future works of openCoT:

\begin{itemize}
\item This version of openCoT only supports CPU allocation. However, the bandwidth allocation plays an essential role too and must be added to openCoT.
\item In this paper, we have tested openCoT using two simple scenarios. More comprehensive benchmarks are needed.
\item Currently, the function execution mechanism of openCoT supports Python objects which can be transformed into JSON strings. In order to provide better performance, a bytes level data transfer should be utilized.
\item In this version of openCoT, there are no security considerations. For commercial use or for more precise academic analysis, considering security protocols (i.g. Node authentication) is vital.
\end{itemize}

\ifCLASSOPTIONcaptionsoff
  \newpage
\fi

\bibliographystyle{IEEEtran}  
\bibliography{referes}

\begin{thebibliography}{10}
\providecommand{\url}[1]{#1}
\csname url@samestyle\endcsname
\providecommand{\newblock}{\relax}
\providecommand{\bibinfo}[2]{#2}
\providecommand{\BIBentrySTDinterwordspacing}{\spaceskip=0pt\relax}
\providecommand{\BIBentryALTinterwordstretchfactor}{4}
\providecommand{\BIBentryALTinterwordspacing}{\spaceskip=\fontdimen2\font plus
\BIBentryALTinterwordstretchfactor\fontdimen3\font minus
  \fontdimen4\font\relax}
\providecommand{\BIBforeignlanguage}[2]{{%
\expandafter\ifx\csname l@#1\endcsname\relax
\typeout{** WARNING: IEEEtran.bst: No hyphenation pattern has been}%
\typeout{** loaded for the language `#1'. Using the pattern for}%
\typeout{** the default language instead.}%
\else
\language=\csname l@#1\endcsname
\fi
#2}}
\providecommand{\BIBdecl}{\relax}
\BIBdecl

\bibitem{cloud:nist}
P.~Mell, T.~Grance \emph{et~al.}, ``The nist definition of cloud computing,''
  2011.

\bibitem{faas:usingFaaS}
A.~Abrahamsson, ``Using function as a service for dynamic application scaling
  in the cloud,'' 2018.

\bibitem{microservices:monolithic}
S.~Daya, N.~Van~Duy, K.~Eati, C.~M. Ferreira, D.~Glozic, V.~Gucer, M.~Gupta,
  S.~Joshi, V.~Lampkin, M.~Martins \emph{et~al.}, \emph{Microservices from
  Theory to Practice: Creating Applications in IBM Bluemix Using the
  Microservices Approach}.\hskip 1em plus 0.5em minus 0.4em\relax IBM Redbooks,
  2016.

\bibitem{microservices:soa}
P.~Di~Francesco, I.~Malavolta, and P.~Lago, ``Research on architecting
  microservices: trends, focus, and potential for industrial adoption,'' in
  \emph{Software Architecture (ICSA), 2017 IEEE International Conference
  on}.\hskip 1em plus 0.5em minus 0.4em\relax IEEE, 2017, pp. 21--30.

\bibitem{faas:sfaas}
F.~Alder, N.~Asokan, A.~Kurnikov, A.~Paverd, and M.~Steiner, ``S-faas:
  Trustworthy and accountable function-as-a-service using intel sgx,''
  \emph{arXiv preprint arXiv:1810.06080}, 2018.

\bibitem{faas:serverlessIsMore}
E.~Van~Eyk, L.~Toader, S.~Talluri, L.~Versluis, A.~Uț{\u{a}}, and A.~Iosup,
  ``Serverless is more: From paas to present cloud computing,'' \emph{IEEE
  Internet Computing}, vol.~22, no.~5, pp. 8--17, 2018.

\bibitem{serverless:QModel}
G.~McGrath and P.~R. Brenner, ``Serverless computing: Design, implementation,
  and performance,'' in \emph{Distributed Computing Systems Workshops (ICDCSW),
  2017 IEEE 37th International Conference on}.\hskip 1em plus 0.5em minus
  0.4em\relax IEEE, 2017, pp. 405--410.

\bibitem{serverless:application:disaster}
T.~Asghar, S.~Rasool, M.~Iqbal, Z.~Qayyum, A.~Noor~Mian, and G.~Ubakanma,
  ``Feasibility of serverless cloud services for disaster management
  information systems,'' 2018.

\bibitem{faas:ubench}
T.~Back and V.~Andrikopoulos, ``Using a microbenchmark to compare function as a
  service solutions,'' in \emph{European Conference on Service-Oriented and
  Cloud Computing}.\hskip 1em plus 0.5em minus 0.4em\relax Springer, 2018, pp.
  146--160.

\bibitem{docker}
D.~Merkel, ``Docker: lightweight linux containers for consistent development
  and deployment,'' \emph{Linux Journal}, vol. 2014, no. 239, p.~2, 2014.

\bibitem{python}
M.~F. Sanner \emph{et~al.}, ``Python: a programming language for software
  integration and development,'' \emph{J Mol Graph Model}, vol.~17, no.~1, pp.
  57--61, 1999.

\bibitem{zmq}
P.~Hintjens, \emph{ZeroMQ: messaging for many applications}.\hskip 1em plus
  0.5em minus 0.4em\relax " O'Reilly Media, Inc.", 2013.

\bibitem{numpy}
N.~Developers, ``Numpy,'' \emph{NumPy Numpy. Scipy Developers}, 2013.

\end{thebibliography}

\end{document}